\documentclass[%
 twocolumn,
 aps,
 amsmath,amssymb,
 prc,longbibliography,
 superscriptaddress,
]{revtex4-2}

\usepackage{graphicx}
\usepackage{dcolumn}
\usepackage{bm}
\usepackage{physics} 
\newcolumntype{d}[1]{D{.}{.}{#1}} 
\usepackage{xcolor}
\usepackage{soul}
\usepackage{amsthm}

\usepackage{orcidlink}
\usepackage{algorithm}
\usepackage{algorithmic}
\usepackage{booktabs}
\usepackage{multirow}
\usepackage{array}

\begin{document}

\title{Emulator-enabled calibration and uncertainty quantification of \texorpdfstring{$\beta$}{beta}-decay calculations within the energy density functional theory}

\author{Y. Saito\,\orcidlink{0000-0003-1320-8903}}
\email{saitoy@frib.msu.edu}
\affiliation{Facility for Rare Isotope Beams, Michigan State University, East Lansing, Michigan 48824, USA}
\affiliation{Department of Physics and Astronomy, University of Tennessee, Knoxville, Tennessee 37996, USA}

\author{A. Ravli\'c\,\orcidlink{0000-0001-9639-5382}}
\email{ravlic@frib.msu.edu}
\affiliation{Facility for Rare Isotope Beams, Michigan State University, East Lansing, Michigan 48824, USA}
\affiliation{Department of Physics, Faculty of Science, University of Zagreb, Bijeni\v cka c. 32, 10000 Zagreb, Croatia}

\author{W. Nazarewicz\,\orcidlink{0000-0002-8084-7425}}
\email{witek@frib.msu.edu}
\affiliation{Facility for Rare Isotope Beams, Michigan State University, East Lansing, Michigan 48824, USA}
\affiliation{Department of Physics and Astronomy, Michigan State University, East Lansing, Michigan 48824, USA}

\date{\today}

\begin{abstract}
Predictions of $\beta$-decay half-lives of neutron-rich nuclei and their uncertainties critically impact our understanding of the astrophysical rapid neutron capture process.
To provide quantified predictions within nuclear energy density functional theory, as customarily done, 
we perform Bayesian calibrations of two coupling constants: the isoscalar proton-neutron pairing strength and the isovector spin-density coupling. 
We apply an E-optimal experimental design for choosing the calibration sets.
We employ the proton-neutron finite-amplitude method with Skyrme energy density functionals using emulators of half-lives and Gamow--Teller resonance energies.
The calculated $Q_\beta$ values are fine-tuned to experimental values.
The contributions to the half-life uncertainty from the theory error,  $Q_\beta$ values, and the parameter uncertainty are all accounted for in the Bayesian framework.
Our predictions for the even-even $N=126$ isotones provide the shortest half-lives for neutron-rich nuclei among the existing theoretical models, which would affect the flow of heavy-element nucleosynthesis. The proposed framework allows quantified theoretical uncertainties to be propagated to astrophysical observables such as abundance patterns and electromagnetic emission.
\end{abstract}

\maketitle

\section{Introduction \label{sec:introduction}}
The importance of $\beta^-$ decay in the astrophysical rapid neutron capture process ($r$-process) has been known since the seminal work on the origin of elements \cite{BBFH, Cameron}.
In general, $\beta$ decays regulate the flow of nucleosynthesis to larger proton number $Z$: $\beta^-$-decay  half-lives leave imprints on the resulting abundance pattern and determine the timescale of the synthesis of heavy nuclei \cite{Langanke2003review, Cowan2021}.
At the neutron shell closures $N=82$ and 126, the $r$-process path moves closer to the stability line.
This leads to the accumulation of abundances and the formation of second and third $r$-process peaks at $A\sim130$ and 195, respectively.
Stimulated by experimental progress at isotope beam facilities, the $\beta$-decay properties of $N=126$ nuclei have gained attention in theoretical studies~\cite{Engel1999,Suzuki2012, Fang2013, Marketin_RQRPA_2016,Horowitz2019, Robin2024, RSN2026}, primarily due to the relevance to the third $r$-process peak.
The rare-earth peak at $A\sim160$ forms near freeze-out, when the material decays back to stability, through the interplay between $\beta$ decay, neutron capture, and nuclear deformations \cite{Surman1997, Mumpower_rep_2012a, Mumpower_rep_2012b, Arcones_rep_2011a}, while fission may also have a significant impact \cite{Eichler2015, Goriely2013}.
Fission processes, including neutron-induced fission and $\beta$-delayed fission in addition to spontaneous fission, determine the endpoint of the $r$-process path, redistributing the abundances of heavy nuclei to lighter regions \cite{Panov2003, Martinezpinedo2007, Panov2010, Eichler2015, Giuliani2018, Mumpower2018, Vassh2019, Giuliani2020}.
As $\beta$ decay determines the timescale for populating fissile nuclei, the contribution of fission may also be affected by trends in $\beta$-decay half-lives.

For $r$-process nucleosynthesis calculations, scalable nuclear $\beta$-decay models capable of computing the decay properties across the chart of nuclides are necessary.
Global theoretical descriptions of $\beta$ decay have seen significant progress since the pioneering work on the gross theory \cite{Takahashi1969, Takahashi1973} and its refinements 
\cite{Nakata1997}. A more microscopic description was offered by the finite-range droplet model (FRDM) accompanied by a quasiparticle random-phase approximation (QRPA)~\cite{Moller1997}, later including first-forbidden (FF) transitions~\cite{Moller2003, Moller2019}.
In recent decades, there have also been remarkable advances in energy density functional (EDF) theory as a scalable many-body method, combined with QRPA, for computing global $\beta$-decay properties.
Initially, the calculations were limited to spherical nuclei \cite{Engel1999, Marketin_RQRPA_2016, Niu2013, Borzov2020}.
The development of the linear-response formulation of charge-changing QRPA~\cite{Mustonen2014,Ravlic2024} enabled global calculations with axial symmetry using nonrelativistic Skyrme EDFs~\cite{MustonenEngel,Ney2020} and relativistic EDFs~\cite{RSN2026}.
The sensitivity of $r$-process simulations to the global trend of $\beta$-decay half-lives has been investigated extensively~\cite {Niu2013, Mumpower2014, Lund2023, Kullmann2023, Hao2023, RSN2026}, given their significant impact on observables such as abundance patterns and signatures of electromagnetic emission.

In our recent study~\cite{RSN2026}, we found that the $\beta$-decay half-lives of nuclei beyond $N=126$ calculated with axially deformed EDF theories agree well among themselves but are significantly longer than those in previous global tabulations, such as Refs.~\cite{Moller2019, Marketin_RQRPA_2016}.
The longer half-lives would slow the production of third $r$-process peak elements and heavy nuclei that undergo neutron-induced fission~\cite{Panov2005, Martinezpinedo2007, Petermann2012}.
However, only a single global tabulation of $\beta$-decay half-lives has existed for nonrelativistic Skyrme EDFs~\cite{Ney2020}, preventing systematic comparison with relativistic EDFs.
Furthermore, the lack of uncertainty on the predictions makes the departure from the earlier global tabulations inconclusive.

In this work, we address these problems by performing Bayesian calibrations of parameters relevant to $\beta$-decay calculations for multiple Skyrme EDFs employing the recently developed emulator~\cite{SMLR}.
This paper is organized as follows.
In Sec.~\ref{sec:methods}, we briefly review the aspects of Skyrme EDF relevant to $\beta$-decay calculations.
We then introduce a correction to theoretical estimates of $Q_\beta$ values.
The framework for Bayesian calibration and uncertainty quantification is discussed in the subsequent subsections, including the emulator architecture and the approach for identifying experimental data that are informative for the calibration.
Section~\ref{sec:results} presents the performance of the emulators,  results of the Bayesian calibration, and
the uncertainty quantification for the even-even $N=126$ isotones.
The conclusions and outlook are given in Sec.~\ref{sec:conclusions}.

\section{Theoretical Methods \label{sec:methods}}
\subsection{\texorpdfstring{$\beta$}{Beta}-decay calculations with the EDF theory \label{subsec:betaEDF}}
Our $\beta$-decay calculations utilize the framework developed in Refs.~\cite{MustonenEngel, Ney2020}.
Within the nuclear EDF theory, the ground state of a nucleus undergoing $\beta$ decay is obtained from the Hartree--Fock--Bogoliubov (HFB) calculations using the \textsc{HFBTHO}~\cite{Perez2017} solver, assuming
axially-deformed and reflection-symmetric nuclear shapes.
Once the ground state is computed, proton-neutron finite-amplitude method (pnFAM)~\cite{Mustonen2014} calculations are performed to obtain the Gamow--Teller (GT) strength functions and $\beta$-decay half-lives. In this work, we use a deformed harmonic oscillator basis with 18 shells. 
The GT strength functions were computed with a smearing half-width of 0.25~MeV, in steps of 0.15~MeV on the real energy axis, with convergence accuracy of $10^{-5}$.  For half-lives, the circular FAM contour was discretized with 200 Gauss--Legendre points, calculated with convergence accuracy of $10^{-9}$.

The calibration of nuclear EDFs is typically performed using the ground-state properties of even-even nuclei, which only require time-even components of the EDFs.
However, $\beta$-decay calculations also require the time-odd terms. For Skyrme EDFs, these are
\begin{align}
    \mathcal{H}^{\text{odd}}_{t\, t_3} (\bm{r}) \equiv & C^{s}_t [\rho_{00}]\, \bm{s}^2_{t\,t_3} + C^{\Delta s}_{t}\; \bm{s}_{t\,t_3} \cdot \bm{\nabla}^2\bm{s}_{t\,t_3} + C^{j}_t \, \bm{j}^2_{t\,t_3} \nonumber\\
    & + C^{T}_t \, \bm{s}_{t\,t_3}\cdot \bm{T}_{t\,t_3} + C^{s \bm{\nabla} j}_t \, \bm{s}_{t\,t_3} \cdot \bm{\nabla} \times \bm{j}_{t\,t_3} \nonumber \\
    & + C^F_t\, \bm{s}_{t\,t_3} \cdot \bm{F}_{t\,t_3} + C^{\bm{\nabla} s}_t\, (\bm{\nabla} \cdot \bm{s}_{t\,t_3})^2, \label{eq:time_odd}
\end{align}
where $t$ and $t_3$ are isospin and its projection, respectively; $\bm{s}_{t\,t_3}$, $\bm{j}_{t\,t_3}$, $\bm{T}_{t\,t_3}$ and $\bm{F}_{t\,t_3}$ are the spin density, current density, spin-kinetic density, and tensor-kinetic density, respectively; and the $C$'s denote the coupling constants.
In $\beta$-decay calculations, the subset of these coupling constants of interest is $\{C_1^s, C_1^T, C_1^F\}$, which are related to the Landau--Migdal parameters~\cite{Bender2002}.
In addition, the isoscalar pairing strength $V_0$ enters the charge-changing response but is not constrained by the ground states of even-even nuclei.
Following Ref.~\cite{MustonenEngel}, which obtained its most robust fits by varying only $\{C_1^s, V_0\}$, we vary these two parameters and fix the remaining time-odd couplings.

In the global $\beta$-decay half-life calculations performed in Ref.~\cite{Ney2020}, the Skyrme EDF SkO$'$~\cite{Reinhard1999} was used.
For a more systematic investigation of the dependence of the $\beta$-decay properties on the EDF, we also consider Skyrme EDFs  SLy4~\cite{Chabanat1998} and UNEDF1-HFB ~\cite{Schunck2015}.

For each functional, the density-dependent part of $C_1^s$ and the coupling $C_1^{\Delta s}$ are set to zero. 
The couplings $C_1^T$ and $C_1^F$ are fixed by local gauge invariance, which relates them to the coupling $C_1^J$ of the time-even $\bm{J}^2$ term, where $\bm{J}$ is the spin-current density; see Refs.~\cite{Perlinska2004,Bender2002} for details.
As a result, $C_1^T=0$ for SLy4 and UNEDF1-HFB, which both omit the $\bm{J}^2$ term, while SkO$'$ retains it and has $C_1^T=-9.2$~MeV\,fm$^5$.
The coupling $C_1^F$ vanishes for all three functionals, as none includes a tensor term.

\subsection{Finetuning of  \texorpdfstring{$Q_\beta$}{Q} values \label{subsec:qcorr}}
In this study, we follow Ref.~\cite{Ney2020} to approximate the $Q_\beta$ values using the HFB input.
Specifically, the $Q_\beta$ values are computed as
\begin{equation}
    Q^{\mathrm{HFB}}_\beta \simeq \Delta M_{n-\mathrm{H}} + \lambda_n - \lambda_p - (E_{n,\mathrm{min}} + E_{p,\mathrm{min}}),
\end{equation}
where $\Delta M_{n-\mathrm{H}}$ is the mass difference between a neutron and a hydrogen atom, $\lambda_{n}$ and $\lambda_p$ are the neutron and proton chemical potentials, respectively, and $E_{n,\mathrm{min}}$ and $E_{p,\mathrm{min}}$ are the lowest neutron and proton quasiparticle energies, respectively.

Existing global $\beta$-decay calculations with a Skyrme EDF employed SkO$^\prime$, partly due to its relative accuracy in predicting $Q_\beta$ values compared to other Skyrme EDFs~\cite{MustonenEngel}.
Since the $\beta$-decay rates are roughly proportional to $Q_{\beta}^5$, a systematic deviation in the calculated $Q_\beta$ values leads to a systematic deviation in the half-lives.
Since the calculated $Q_\beta$ values of even-even nuclei do not depend on $V_0$ and $C_1^s$,
the calibration of these parameters cannot remove such a deviation.
For these reasons, in this work, we apply a linear correction to the calculated $Q_\beta$ values for each functional and treat the discrepancy from the mean value as a statistical error.

\begin{figure}[t]
\includegraphics[width=\columnwidth]{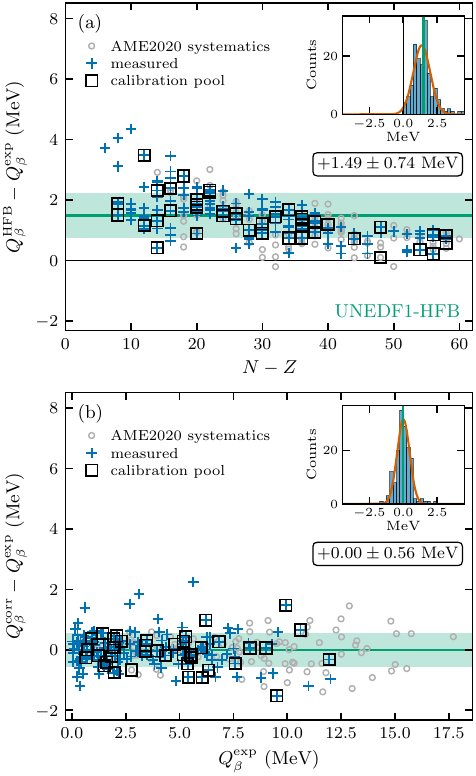}
\caption{\label{fig:q_correction}%
Deviation of the calculated $Q_\beta$ values from experiment for UNEDF1-HFB: (a) before the correction as a function of $N-Z$; (b) after the correction as a function of $Q^{\rm exp}_\beta$.
Crosses mark the nuclei with measured $Q_\beta$ used in the fit, gray circles indicate the nuclei whose masses are based on systematics in AME2020 (not used in the fit), and squares are the nuclei in Table~\ref{tab:pool_hl}.
The horizontal line and shaded band show the mean and statistical 
deviation.
The insets show the distribution of residuals with a normal fit.
}
\end{figure}

HFB calculations were performed to obtain $Q^{\mathrm{HFB}}_\beta$ for even-even nuclei with $18 \leq Z \leq 96$ and a positive value of  $Q^\mathrm{exp}_\beta$ for $\beta^-$ decays listed in AME2020~\cite{AME2020}.
The corrected theoretical $Q_\beta$ values are defined as:
\begin{equation}
    Q^\mathrm{corr}_\beta = Q^\mathrm{HFB}_\beta - \left[a_0 + a_T (N-Z) +  a_A A \right], \label{eq:qcorr}
\end{equation}
where the coefficients $a_0$, $a_T$, and $a_A$ were obtained by an unweighted least-squares fit of $Q^\mathrm{HFB}_\beta-Q^\mathrm{exp}_\beta$ to the 154 nuclei with a measured $Q_\beta$.
The $Q_\beta$ values from systematics in AME2020 were not used in the fit.
Figure~\ref{fig:q_correction} shows the deviation of the theoretical $Q_\beta$ values from experiment computed with the UNEDF1-HFB functional, before and after the correction.
The fit results are summarized in Table~\ref{tab:qcorr}.
We found that a linear correction in the directions of isospin and $A$ yields approximately normally distributed residuals.
After the correction, the residuals exceed $2\sigma_Q$ in magnitude in a few regions.
The isotopes $^{42,44,46}$Ar are overestimated by 2--4$\sigma_Q$ and the neutron-rich Ti, Cr, and Fe isotopes are underestimated by up to $3\sigma_Q$ for all three functionals.
Nuclei at or near shell closures, such as $^{48}$Ca, $^{68}$Ni, $^{80}$Zn, $^{80}$Ge, and $^{126,132}$Sn, also exceed $2\sigma_Q$ with at least one functional.
This suggests that the errors in the description of masses and shell structure near magic numbers have local features that smooth corrections cannot
accommodate.

\begin{table}[t]
\caption{\label{tab:qcorr}%
Summary of the linear correction of the calculated $Q_\beta$ values, Eq.~(\ref{eq:qcorr}), for each functional:
mean and standard deviation of $Q^{\rm HFB}_\beta-Q^{\rm exp}_\beta$ before the correction, the fitted coefficients, and the residual standard deviation $\sigma_Q$ after the correction, over the 154 even-even nuclei with a measured $Q_\beta$ in AME2020~\cite{AME2020}. The mean after correction is zero. All values in MeV.}
\begin{ruledtabular}
\begin{tabular}{lccccc}
Functional & before & $a_0$ & $a_T$ & $a_A$ & $\sigma_Q$ \\
\colrule
SkO$'$ & $+0.22 \pm 0.69$ & $+0.789$ & $-0.0752$ & $+0.0122$ & 0.58 \\
SLy4 & $+0.17 \pm 0.63$ & $+0.080$ & $-0.0376$ & $+0.0091$ & 0.62 \\
UNEDF1-HFB & $+1.49 \pm 0.74$ & $+2.427$ & $-0.0590$ & $+0.0057$ & 0.56 \\
\end{tabular}
\end{ruledtabular}
\end{table}

The corrected $Q^{\mathrm{corr}}_\beta$ values are then used in the pnFAM half-life calculations, with the integration contours and the phase-space factors also adjusted to the corrected values.
This means that the upper limit of the integration contour, the maximum QRPA energy $E_{\max}^{\rm QRPA}$, is defined as $E_{\max}^{\rm QRPA} = Q_\beta^\mathrm{corr} + E_{n,\mathrm{min}} + E_{p,\mathrm{min}}$, and the electron endpoint energy that enters the phase-space factors is shifted by $Q^{\mathrm{corr}}_\beta - Q^{\mathrm{HFB}}_\beta$, while the strength itself is unchanged.
The standard deviation of the residuals after the correction, $\sigma_Q$ (Table~\ref{tab:qcorr}), is propagated to the $\beta$-decay half-lives as discussed in Sec.~\ref{subsec:bayes}.

\subsection{Calibration parameters and observables\label{subsec:observables}}
The two parameters calibrated in this work are $\bm{\theta}=(V_0, C_1^s)$, i.e., the isoscalar proton-neutron pairing strength $V_0$ and the isovector spin density coupling constant $C_1^s$.
Both are in MeV\,fm$^3$.
Initial samples are generated over the intervals  $V_0\in[-500,0]$\,MeV\,fm$^3$ and $C_1^s\in[0,250]$\,MeV\,fm$^3$  (Sec.~\ref{subsec:training}), and throughout the paper we also use the dimensionless parameters
\begin{gather}
    v=\frac{V_0+250}{250},\nonumber\\
    c=\frac{C_1^s-125}{125},\nonumber\\
    \bm{z}\equiv(v,c)\in[-1,1]^2,
    \label{eq:scaled_params}
\end{gather}
which map the ranges to $[-1,1]$.

We note, however, that at sufficiently strong isoscalar pairing, a proton-neutron pairing phase transition is known to occur~\cite{Engel1997}.
In the current work, the valid range of the parameters is restricted to the region in which all the nuclei in the calibration (see Table~\ref{tab:pool_hl} and Sec.~\ref{subsec:dataset}) are stable against the phase transition, $V_0 \geq V_{\min}$ with $V_{\min}=-371$~MeV\,fm$^3$ for SkO$'$, $-450$~MeV\,fm$^3$ for SLy4, and $-360$~MeV\,fm$^3$ for UNEDF1-HFB.
We also require $C_1^s \geq 50$~MeV\,fm$^3$, since below this value half-life calculations become unstable.
We refer to this region as the stable domain.

The first observable is the total $\beta^-$-decay half-life, which includes allowed GT and FF transitions, computed as in Ref.~\cite{Ney2020}.
Because the calculated half-lives vary over many orders of magnitude across the sampled parameter space, we instead consider $\log_{10}(T_{1/2}/{\rm s})$.

The second observable is the GT resonance energy.
The GT strength function is computed as a function of the QRPA energy $E^\mathrm{QRPA}$, which is then converted to the excitation energy of the daughter nucleus, $E_x = E^\mathrm{QRPA} - (E_{n,\mathrm{min}} + E_{p,\mathrm{min}})$, approximating the daughter ground state by the lowest two-quasiparticle state as in Sec.~\ref{subsec:qcorr}~\cite{Engel1999}.

For the GT resonance, we focus on the strength around the experimental resonance energies.
Based on the GT resonance spectra in the literature (Table~\ref{tab:pool_gt}), we define a 6~MeV window around the observed resonance peak, within which we compute the centroid of the calculated GT strength.
For $^{48}$Ca, the window is $[4.5, 14.5]$~MeV, based on Ref.~\cite{Ca48GT}.

To avoid a sharp cutoff in energy when computing the centroid within the energy window, we use a smooth window function.
For a window $[E_{\rm lo}, E_{\rm hi}]$, the weight assigned to strength at the excitation energy $E_x$ is
\begin{equation}
w(E_x)
=
\left[1+e^{-(E_x-E_{\rm lo})/\tau}\right]^{-1}
\left[1+e^{-(E_{\rm hi}-E_x)/\tau}\right]^{-1},
\end{equation}
where $\tau$ controls the smoothing width.
In this work we use $\tau = 0.5~{\rm MeV}$.
In the limit $\tau \rightarrow 0$, the smooth window approaches a sharp window.
The  strength centroid is then computed as
\begin{equation}
\bar{E}
=
\frac{\int dE_x\, E_x\, w(E_x) S(E_x)}
     {\int dE_x\, w(E_x) S(E_x)}.
\end{equation}
In the calibration, $\bar{E}$ is compared with the measured peak energy $E_{\rm GT}^{\rm exp}$.
The smooth cutoff makes the centroid continuous with respect to changes in the predicted GT strength distribution, which improves numerical stability during emulator training and Bayesian calibration.
We use the centroid rather than the location of the highest peak, because in a fragmented strength distribution the highest peak can jump from one state to another as parameters vary.

\subsection{Emulator architecture\label{subsec:emulator}}
The emulators used in this work are based on the reduced-order model developed in Ref.~\cite{SMLR}.
It models the dependence of a quantity of interest on the calibration parameters through the eigendecomposition of a small matrix.
Each emulator predicts one observable, a half-life or a GT resonance centroid, and a separate emulator is trained for each nucleus, each observable, and each functional.

\subsubsection{Trainable matrix}
The starting point of the emulator is a trainable matrix in terms of the scaled parameter $\bm{z}=(v,c)$, defined as
\begin{equation}
    \hat{\mathcal{M}}(\bm{z}) = \mathcal{M}_0
    + (v - v_0)\,\mathcal{M}_1 + (c - c_0)\,\mathcal{M}_2,
    \label{eq:PMM}
\end{equation}
where $\mathcal{M}_0$ is a real-valued diagonal $n\times n$ matrix, $\mathcal{M}_1$ and $\mathcal{M}_2$ are real symmetric $n\times n$ matrices, and $\bm{z}_0=(v_0,c_0)$ is a training data point, closest to the mean.
The elements of $\mathcal{M}_0$, $\mathcal{M}_1$, and $\mathcal{M}_2$ are $n+n(n+1)$ trainable parameters.
Evaluating the emulator prediction at a parameter point $\bm{z}$ amounts to building $\hat{\mathcal{M}}(\bm{z})$ and diagonalizing it,
\begin{equation}
    \hat{\mathcal{M}}(\bm{z})\,\bm{u}^{(i)}(\bm{z})
    = \hat{m}^{(i)}(\bm{z})\,\bm{u}^{(i)}(\bm{z}),
    \qquad i=1,\dots,n,
\end{equation}
with the eigenvalues ordered as $\hat{m}^{(1)}\le\hat{m}^{(2)}\le\dots\le\hat{m}^{(n)}$.
Since the matrix depends linearly on the parameters, its eigenvalues and eigenvectors are smooth functions of $\bm{z}$ except at eigenvalue crossings, and training determines how these functions depend on the parameters.

\subsubsection{Scalar observable emulator}
The scalar emulator predicts the observable directly as the middle eigenvalue of Eq.~(\ref{eq:PMM})~\cite{SMLR}.
We use a matrix of dimension $n=5$, so that the observable is emulated by the third eigenvalue,
\begin{equation}
    \hat{y}(\bm{z}) = \hat{m}^{(3)}(\bm{z}),
\end{equation}
where $\hat{y}$ is $\log_{10}(T_{1/2}/{\rm s})$ or $\bar{E}$.
Each scalar emulator has $n+n(n+1)=35$ trainable parameters.

\subsection{Training of the emulators \label{subsec:training}}
For each functional, 100 parameter sets $\bm{\theta}=(V_0, C_1^s)$ were drawn from a two-dimensional Sobol quasi-random sequence~\cite{Sobol1976}.
pnFAM calculations were performed at each point to obtain $\beta$-decay half-lives and GT strength functions up to a QRPA energy of 30~MeV.
Samples outside the stable domain were discarded from the half-life training data, as were samples that did not converge or with decay rates below $10^{-9}$~s$^{-1}$, which are dominated by numerical noise.

The emulators are trained by minimizing the mean squared error of the prediction, $\log_{10}(T_{1/2}/{\rm s})$ or $\bar{E}$, relative to the original pnFAM calculations over the training samples.

All emulators are trained with the Adam gradient descent optimizer~\cite{Kingma2017} in \texttt{TensorFlow}~\cite{tensorflow2015} for 30\,000 iterations with an initial learning rate of 0.01, using four different random seeds and keeping the best result.

Emulator accuracy was evaluated by five-fold cross-validation.
The samples for each training target were split into five parts, and each fold used a different 80\% of the samples for training while holding out the remaining 20\% for validation.
For half-lives, the samples were split so that every fold roughly spans the full range of half-lives.
For each training target, the emulator used in the calibration is taken from the fold with the lowest validation loss, trained on 80\% of the samples.
The emulator error $\sigma_{\rm em}$ in Eq.~\eqref{eq:variance_decomposition} was estimated by pooling the residual of the held-out predictions across the five folds.
This estimate is used in the preliminary Bayesian calibration described in Sec.~\ref{subsec:calib_optimal_sets}.
For the half-lives, the final value of $\sigma_{\rm em}$ is estimated from separate test calculations in the posterior region of the preliminary calibration (Sec.~\ref{subsec:calib_optimal_sets}).
For the GT resonance centroids, the cross-validation estimate is retained.

\subsection{Bayesian inference\label{subsec:bayes}}
The calibration targets are the experimentally measured half-lives from NUBASE2020~\cite{Kondev2021} and GT resonance peak energies of a set of nuclei.
The quantities of interest are the two parameters $\bm{\theta}$.
The scales of the theory error, $\sigma^{\rm HL}_{\rm th}$ and $\sigma^{\rm GT}_{\rm th}$, are inferred at the same time.
The posterior is expressed as
\begin{gather}
    p(\bm{\theta},\sigma^{\rm HL}_{\rm th},\sigma^{\rm GT}_{\rm th}
      \mid \mathcal{D})\nonumber \\
    \propto
    \pi(\bm{\theta})\,\pi(\sigma^{\rm HL}_{\rm th})\,
    \pi(\sigma^{\rm GT}_{\rm th})
    \prod_{k\in\mathcal{C}_{\rm HL}}\mathcal{L}_k
    \prod_{k\in\mathcal{C}_{\rm GT}}\mathcal{L}_k ,
    \label{eq:posterior}
\end{gather}
where $\mathcal{D}$ is the calibration dataset, $\mathcal{C}_{\rm HL}$ and $\mathcal{C}_{\rm GT}$ are the sets of nuclei included in the calibration with measured half-lives and GT resonance peak energies, respectively, $\pi$ denotes the priors, and $\mathcal{L}_k$ the likelihood of nucleus $k$ in the calibration set.

Each measured calibration target is compared with the corresponding emulator prediction $\hat{y}_k(\bm{\theta})$ through a likelihood $\mathcal{L}_k$.
We use a normal likelihood and a Student-$t$ likelihood with $\nu = 4$ degrees of freedom, a typical choice for robust inference~\cite{Lange1989,Gelman2013BDA}, to check the sensitivity of the results to outliers.
With its heavier tail, the Student-$t$ likelihood reduces the weight of the few nuclei whose half-lives the prediction misses by several orders of magnitude, and it is used for the main results.
For both likelihoods, the variance is decomposed as
\begin{gather}
    \sigma_{{\rm tot},k}^2 = \sigma_{\rm th}^2
      + \sigma_{{\rm exp},k}^2 + \sigma_{{\rm em},k}^2 + \sigma_{Q_\beta,k}^2 ,
    \label{eq:variance_decomposition}
\end{gather}
with $\sigma_{\rm th}=\sigma^{\rm HL}_{\rm th}$ for half-lives and $\sigma^{\rm GT}_{\rm th}$ for GT resonance centroids.
The four components in Eq.~(\ref{eq:variance_decomposition}) are the theory error, the experimental uncertainty, the emulator error, and the error propagated from the uncertainty of the corrected $Q_\beta$ value.
For the Student-$t$ likelihood, the scale parameter is set to $\sigma_{{\rm tot},k}\sqrt{(\nu-2)/\nu}$, so that its variance equals $\sigma_{{\rm tot},k}^2$ and the two likelihoods differ only in the shape of their tails.
The reported experimental uncertainty for half-life $\Delta T_{1/2}$ is converted to $\sigma_{\rm exp}=\Delta T_{1/2}/(T_{1/2}\ln 10)$.
The $\sigma_{Q_\beta,k}^2$ term only enters the half-life variance, and it is zero for the GT resonance centroid.
The propagation of the standard deviation of $Q_\beta$ values assumes that $\beta$-decay rates are proportional to $Q_\beta^5$, resulting in $\sigma_{Q_\beta,k}=5\sigma_Q/(Q^{\rm corr}_{\beta,k}\ln 10)$ in units of $\log_{10}T_{1/2}$, where $Q^{\rm corr}_{\beta,k}$ is the corrected $Q_\beta$ value for the nucleus $k$ and $\sigma_Q$ is the standard deviation of the fit residual obtained in Sec.~\ref{subsec:qcorr}.
For GT resonance centroids, $\sigma_{\rm exp}$ is the uncertainty of the measured resonance energy in MeV.
The emulator error is fixed per nucleus as described in Sec.~\ref{subsec:training}.

The shared theory error $\sigma_{\rm th}$, rather than a nucleus-dependent error, is the natural choice.
Each observable of a nucleus, half-life or GT resonance energy, has a single measured value, so a nucleus-dependent error would simply absorb the residual of the theoretical prediction, whereas the shared error represents the spread of the residuals over the calibration set.
$\sigma_{\rm th}$ measures how far the functional, at its optimal parameter values, is from describing the calibration set as a whole.
In Ref.~\cite{MustonenEngel} the corresponding widths of the $\chi^2$ distributions were set by hand, while in this work they are inferred from the data.

The prior distributions are defined as,
\begin{align}
    V_0     &\sim \mathcal{N}_{[V_{\min},\,0]}(V_1,\,200)~{(\rm in\,MeV\,fm^3)}, \\
    C_1^s   &\sim \mathcal{U}(50,\,250)~{(\rm in\,MeV\,fm^3)}, \\
    \sigma^{\rm HL}_{\rm th},\;\sigma^{\rm GT}_{\rm th}
            &\sim \mathrm{Half}\text{-}t_{\nu=4}(0,\,1),
\end{align}
where $\mathcal{N}_{[a,b]}(\mu,s)$ denotes a normal distribution with the mean $\mu$ and standard deviation $s$, truncated to $[a,b]$.
The truncation at $V_{\min}$ was implemented as a smooth taper, $\tanh^2[(V_0-V_{\min})/w]$, with $w=10$~MeV\,fm$^3$, so the prior density goes to zero continuously at the limit.
$V_1$ denotes the like-particle isovector pairing strength: $-264.2$~MeV\,fm$^3$ for SkO$'$, $-332.7$~MeV\,fm$^3$ for SLy4, and $-247.4$~MeV\,fm$^3$ for UNEDF1-HFB.
$\mathcal{U}(50,\,250)$ is a uniform distribution that spans the range between 50 and 250.
The support of the priors on $\bm{\theta}$ is the region where the emulators were trained (Sec.~\ref{subsec:training}); outside this region, the emulators are not defined.
The sampled range of $C_1^s$, $[0,250]$~MeV\,fm$^3$, was chosen with the reported value for SkO$'$ from Ref.~\cite{MustonenEngel}, 128.279~MeV\,fm$^3$, roughly at its center, and the lower bound of the prior is the stability limit of Sec.~\ref{subsec:observables}.
These priors were designed to be weakly informative, with the largest density at $V_0 = V_1$.
The half-$t$ prior on the size of the theory error was chosen to be nonnegative and heavy-tailed, allowing values well above its scale of 1.

The posterior is sampled with the sequential Monte Carlo (SMC) sampler of \texttt{PyMC}~\cite{pymc}, with four independent chains of 2000 draws each.

The posterior predictive distribution of a half-life for a nucleus $k$ is
\begin{gather}
    p(y_k \mid \mathcal{D})= \nonumber \\=
    \int p\!\left(y_k \mid \bm{\theta}, \sigma^{\rm HL}_{\rm th}\right)\,
    p\!\left(\bm{\theta}, \sigma^{\rm HL}_{\rm th}, \sigma^{\rm GT}_{\rm th} \mid \mathcal{D}\right)\,
    d\bm{\theta}\, d\sigma^{\rm HL}_{\rm th}\, d\sigma^{\rm GT}_{\rm th},
    \label{eq:predictive}
\end{gather}
where $p(y_k \mid \bm{\theta}, \sigma^{\rm HL}_{\rm th})$ is the half-life likelihood with variance $\sigma_{\rm th}^2 + \sigma_{Q_\beta, k}^2$, centered at the model half-life $\hat{y}_k(\bm\theta)$.
Note that $y_k$ denotes the predicted variable, whereas $\hat{y}_k(\bm{\theta})$ is the model half-life at given $\bm{\theta}$.
The experimental and emulator terms of 
Eq.\,(\ref{eq:variance_decomposition}) 
are absent for predictions with no measurement or emulators.
For the nuclei in the calibration pool, we follow the same procedure, using the emulator as $\hat{y}_k(\bm\theta)$ and the full variance in Eq.~(\ref{eq:variance_decomposition}).

In this work, we demonstrate uncertainty quantification with $N=126$ isotones.
Evaluating $\hat{y}_k(\bm{\theta})$ with pnFAM at every posterior sample is costly and impractical.
Instead, for each nucleus, a quadratic surface in $(V_0, C_1^s)$ is fitted to 20 pnFAM calculations at parameter values drawn from the posterior and used as $\hat{y}_k(\bm{\theta})$.
We take this approach since it is significantly simpler than training and validating an emulator for each observable, which would also require more samples.
As discussed in Sec.~\ref{subsec:N126}, this introduces negligible error.
The integral is then evaluated by drawing one value of $y_k$ for every posterior sample.

\subsection{Selection of calibration observables and nuclei\label{subsec:dataset}}

While GT resonance data mostly constrain the spin-density coupling constant $C^s_1$, half-lives are needed to constrain the isoscalar pairing strength $V_0$.
As a pool of candidate experimental data for our calibration, we consider the union of the half-lives and GT resonance measurements listed in Refs.~\cite{MustonenEngel} and \cite{Ravlic2025}.
Additionally, we also considered the half-lives of $^{56}$Cr, $^{208}$Hg ($N=126$ region), and $^{234}$Ra (a deformed nucleus, beyond $N=126$) to cover regions of the chart of nuclides with relatively low $Q_\beta$ values, and $^{132}$Cd for its large $Q_\beta$ value and relevance to the second $r$-process peak.
In total, 42 measured half-lives and 8 GT resonance energies were considered as candidates.
The candidate nuclei are listed in Tables~\ref{tab:pool_hl} and \ref{tab:pool_gt}.

When choosing experimental data for calibration, it is important to select those for which the corresponding theoretical calculations are highly sensitive to parameter variations.
To quantify the sensitivity of the theoretical models to parameter variation, we introduce the information matrix and E-optimal experimental design.

\subsubsection{Information matrix}
For a calibration set $\mathcal{C}$, the information matrix is defined as
\begin{equation}
    \bm{M}_\mathcal{C}(\bm{z}_t) = \sum_{x\in\mathcal{C}}\frac{\bm{f}(x, \bm{z}_t) \bm{f}^T(x, \bm{z}_t)}{\sigma^2(x)} \equiv \sum_{x \in \mathcal{C}}\bm{\mu}(x, \bm{z}_t), \label{eq:mu}
\end{equation}
where $x$ is the label for a nucleus and its type of observables, $\bm{f}(x, \bm{z}_t)$ is the sensitivity of the observable to the scaled parameters $\bm{z}=(v,c)$ at a reference point $\bm{z}_t$, and $\sigma^2(x)$ is the variance of the residual.
The inverse of $\bm{M}_\mathcal{C}$ approximates the covariance matrix of the parameters in the scaled parameter space (Appendix~\ref{app:information_matrix}).
We take $\sigma^2(x)=\sigma_{\rm exp}^2(x)+\sigma_{\rm em}^2(x)+\sigma_{\rm th}^2+\sigma_{Q_\beta}^2(x)$ for half-lives and $\sigma^2(x)=\sigma_{\rm exp}^2(x)+\sigma_{\rm em}^2(x)+\sigma_{\rm th}^2$ for GT resonance energies, see Eq.~\eqref{eq:variance_decomposition}.

Evaluating Eq.~\eqref{eq:mu} requires two inputs that are not available before any calibration has been done: the theory errors $\sigma_{\rm th}$, which are inferred from data, and the reference point $\bm{z}_t$, which should lie in the region of parameter space that the data favor.
Since it is the calibration set that we are selecting, these inputs must come from a calibration set chosen independently of the selection.
We use the set of Ref.~\cite{MustonenEngel}, whose calibration was subsequently used in the global calculation of Ref.~\cite{Ney2020}, and we hereafter refer to it as Set A (Table~\ref{tab:calibsets}).
The emulator errors come from the cross-validation, and the theory errors and reference point come from the posterior of the Set A calibration (Sec.~\ref{subsec:baseline_calibration}).
The selected sets therefore depend on Set A through $\sigma_{\rm th}$ and $\bm{z}_t$.
As discussed below, we test the dependence on $\bm{z}_t$ by also fitting $\bm{f}(x, \bm{z}_t)$ over the entire parameter domain instead of the region favored by Set A.

Although $\bm{f}(x, \bm{z}_t)$ is defined in Eq.~\eqref{eq:derivative} as the pointwise derivative, in numerical implementation, a linear fit by ordinary least squares is used to approximate the derivative.
That is,  for each nucleus and its observable $x$, the function
\begin{equation}
    \eta(x,\bm{z})
    \simeq
    \alpha(x)+b_V(x)v+b_C(x)c, \label{eq:linear}
\end{equation}
is fit to the pnFAM calculations of $x$ over some region in the parameter space, where $\eta(x,\bm{z})$ is the calculated observable ($\log_{10}T_{1/2}$ or GT resonance centroid energy) at $\bm{z}$ (Appendix~\ref{app:information_matrix}).
The sensitivity vector used in the information matrix is then
\begin{equation}
    \bm f(x, \bm{z}_t)\simeq \bigl(b_V(x),b_C(x)\bigr)^T. \label{eq:sensitivity}
\end{equation}
Therefore, these coefficients represent the average sensitivity over the samples included in the fit and approximate the pointwise derivative.

We consider two prescriptions for estimating the sensitivity of the half-lives: ``global'' estimates and ``local'' estimates.
In the global estimation, the slopes in Eq.~\eqref{eq:linear} are estimated over the valid sampled parameter region.
In the local case, the slopes are fitted to the samples with $V_0$ inside the 16--84\% posterior interval of the Set A calibration (Table~\ref{tab:posteriors}) with the normal likelihood, regardless of the values of $C_1^s$.
This region replaces the reference point $\bm{z}_t$ in Eq.~\eqref{eq:taylor}.

For the GT resonance centroid energy, only data on eight nuclei are currently considered, and each datum is highly sensitive to $C_1^s$.
Therefore, all of them are included in the calibration.
Their slopes are estimated over the whole sampled parameter region.

\subsubsection{E-optimality}
The information matrix $\bm{M}_\mathcal{C}$ measures how well the calibration set $\mathcal{C}$ constrains the parameters, and therefore provides the basis for choosing $\mathcal{C}$.
In general, it is not possible to directly find a calibration set where $\bm{M}_\mathcal{C}$ is globally optimal.
Instead, one needs to employ a scalar criterion to find an optimal calibration set in the specific context.
In this work, we employ E-optimality, which avoids leaving any parameter combination weakly constrained \cite{fedorov2013optimal}.
In practice, this means choosing the calibration set $\mathcal{C}$ that maximizes the smallest eigenvalue of the information matrix $\bm{M}_\mathcal{C}$ \cite{Laky2026}.

Since the set of GT resonance nuclei is fixed, its contribution to 
\begin{equation}
    \bm M_{\rm GT}
    =
    \sum_{x\in\mathcal C_{\rm GT}}\bm\mu(x),
\end{equation}
is constant, where $\bm{\mu}(x)$ is as defined in Eq.~\eqref{eq:mu}.
The E-optimal set for half-lives is therefore
\begin{equation}
    \mathcal C_{{\rm HL},m}^{*}
    =
    \underset{
      \substack{
        \mathcal C_{\rm HL}\subseteq\mathcal P_{\rm HL}\\
        |\mathcal C_{\rm HL}|=m
      }}
      {\arg\max}
    \;\lambda_{\min}\left[
        \bm M_{\rm GT}
        +\sum_{x\in\mathcal C_{\rm HL}}\bm\mu(x)
    \right], \label{eq:e-optimal}
\end{equation}
where $\mathcal{C}^*_{{\rm HL},m}$ is a set of nuclei chosen from the pool $\mathcal{P}_{\rm HL}$, and $\lambda_{\min} [\bm{M}]$ is the smallest eigenvalue of the matrix $\bm{M}$.
We set $m=10$ for half-life data.
Since the GT resonance energies predominantly constrain $C^s_1$, the E-optimality improves the constraint along $V_0$.

The E-optimal sets of ten half-lives were found by computing $\lambda_{\min}$ in Eq.~\eqref{eq:e-optimal} for every subset of $m=10$ half-life nuclei from the pool (Table~\ref{tab:pool_hl}), subject to the requirement that nuclei of different mass, $Q_\beta$ value, shell closure, deformation, and FF-decay contribution are all represented.

\subsection{Calibration procedure \label{subsec:calibration_procedure}}
The calibration requires multiple steps of Bayesian inference to construct E-optimal sets, estimate the parameter regions of interest to evaluate the emulator performance, and perform a final Bayesian calibration.
The complete procedure is the following:
\begin{enumerate}
    \item Using Set A, perform Bayesian inference on $\bm{\theta}=(V_0, C_1^s)$ and construct the global and local E-optimal sets from its posterior (Sec.~\ref{subsec:dataset}),
    \item Perform Bayesian inference on $\bm{\theta}$ with $\sigma_{\rm em}$ estimated from the cross-validation, for the local and global E-optimal sets and the Full set of all candidate nuclei,
    \item Define rectangular regions of $\bm{\theta}$ to cover the obtained posterior distributions and draw samples to generate test data with pnFAM,
    \item Perform final error evaluation of the half-life emulators with the test data to obtain updated $\sigma_{\rm em}$, and
    \item Perform Bayesian inference again with the updated (test) emulator error $\sigma_{\rm em}$.
\end{enumerate}

\section{Results and Discussion \label{sec:results}}
\subsection{Performance of the emulators \label{subsec:emulator_performance}}
For each functional, the emulators were trained on the samples that satisfied the criteria described in Sec.~\ref{subsec:training}.
Therefore, the number of usable samples varies by nucleus.
The half-life and GT resonance centroid emulators are those described in Sec.~\ref{subsec:emulator}, implemented in a locally modified version of Ref.~\cite{SMLR_github}.

\begin{figure}[t]
\includegraphics[width=\columnwidth]{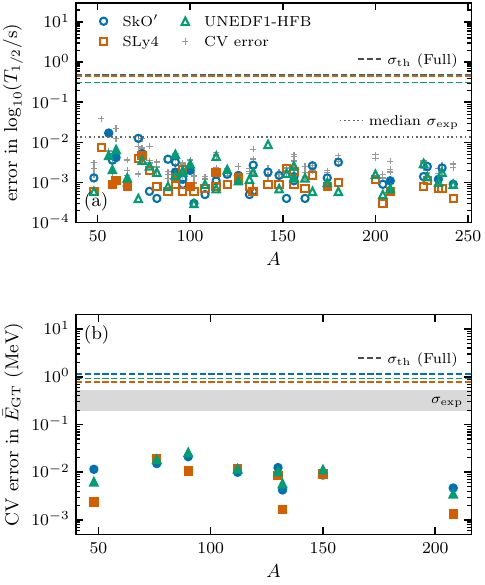}
\caption{\label{fig:emulator_summary}%
Emulator error estimates for each EDF considered in this work.  (a) Half-lives; test errors are marked with colored symbols and cross-validation errors with crosses; (b) GT resonance centroid energies; cross-validation errors are shown in colored symbols, which are retained in the final inference.
Filled symbols indicate members of the L10 sets (identical to the G10 set for UNEDF1-HFB).
Dashed lines are the inferred theory errors from the full-set calibration (Sec.~\ref{subsec:calib_optimal_sets}); the dotted line and the gray band are the experimental uncertainties.}
\end{figure}

Figure~\ref{fig:emulator_summary} shows the emulator errors for each nucleus.
For the GT resonance centroids, the cross-validation errors are one-to-two orders of magnitude smaller than the experimental uncertainties.
The half-life emulators, in contrast, are less accurate, because the half-lives depend on the low-lying strength below the $Q_\beta$ value, which does not necessarily vary smoothly with the calibration parameters.
Their test errors are still mostly below the experimental uncertainties and at most of the same order of magnitude.
For SLy4, we computed 100 additional samples for $^{52}$Ti to reach the desired accuracy.

\begin{table*}[t]
\caption{\label{tab:calibsets}%
Calibration sets, which consist of $\beta$-decay half-life (HL) nuclei and Gamow--Teller
resonance (GT) nuclei. Set A is the literature baseline of Ref.~\cite{MustonenEngel},
common to all functionals. The optimal sets G10 (sensitivities over the full parameter domain)
and L10 (sensitivities evaluated within the Set A posterior interval of $V_0$) are selected independently
for each functional. The Full set contains every pool nucleus of the functional.}
\begin{ruledtabular}
\begin{tabular}{llll}
Functional & Set & Half-life nuclei & GT-resonance nuclei \\
\colrule
  & A & $^{48}$Ar, $^{60}$Cr, $^{72}$Ni, $^{82}$Zn, $^{92}$Kr, $^{102}$Sr, $^{114}$Ru, $^{126}$Cd, & $^{48}$Ca, $^{76}$Ge, $^{90}$Zr, $^{112}$Sn, $^{130}$Te, $^{208}$Pb \\
  &   & $^{134}$Sn, $^{148}$Ba &  \\
SkO$'$ & G10 & $^{56}$Cr, $^{60}$Cr, $^{66}$Fe, $^{74}$Zn, $^{92}$Sr, $^{100}$Zr, $^{114}$Pd, $^{126}$Cd, & all eight nuclei of Table~\ref{tab:pool_gt} \\
  &   & $^{156}$Sm, $^{208}$Hg &  \\
  & L10 & $^{56}$Cr, $^{60}$Cr, $^{66}$Fe, $^{74}$Zn, $^{92}$Sr, $^{100}$Zr, $^{114}$Pd, $^{126}$Cd, & same as G10 \\
  &   & $^{208}$Hg, $^{234}$Ra &  \\
  & Full & all 40 nuclei of Table~\ref{tab:pool_hl} not marked for this & same as G10 \\
  &   & functional &  \\
\colrule
  & A & same as SkO$'$ & same as SkO$'$ \\
SLy4 & G10 & $^{52}$Ti, $^{58}$Ti, $^{60}$Cr, $^{74}$Zn, $^{92}$Sr, $^{100}$Zr, $^{114}$Pd, $^{126}$Cd, & all eight nuclei of Table~\ref{tab:pool_gt} \\
  &   & $^{174}$Er, $^{208}$Hg &  \\
  & L10 & $^{58}$Ti, $^{60}$Cr, $^{66}$Fe, $^{74}$Zn, $^{92}$Sr, $^{100}$Zr, $^{114}$Pd, $^{126}$Cd, & same as G10 \\
  &   & $^{174}$Er, $^{208}$Hg &  \\
  & Full & all 40 nuclei of Table~\ref{tab:pool_hl} not marked for this & same as G10 \\
  &   & functional &  \\
\colrule
  & A & same as SkO$'$ & same as SkO$'$ \\
UNEDF1-HFB & G10 & $^{56}$Cr, $^{58}$Ti, $^{60}$Cr, $^{66}$Fe, $^{92}$Sr, $^{100}$Zr, $^{120}$Pd, $^{126}$Cd, & all eight nuclei of Table~\ref{tab:pool_gt} \\
  &   & $^{156}$Sm, $^{208}$Hg &  \\
  & L10 & same as G10 & same as G10 \\
  & Full & all 40 nuclei of Table~\ref{tab:pool_hl} not marked for this & same as G10 \\
  &   & functional &  \\
\end{tabular}
\end{ruledtabular}
\end{table*}
\subsection{Baseline calibration with the literature set \label{subsec:baseline_calibration}}
To establish a reference calibration, we employ the dataset used in Ref.~\cite{MustonenEngel}, which we refer to as Set A (see Table~\ref{tab:calibsets}).
Here we perform Bayesian inference using Set A with the normal likelihood of Sec.~\ref{subsec:bayes}, which is the closest to the $\chi^2$ fit of Ref.~\cite{MustonenEngel}.
The results with the Student-$t$ likelihood are listed in Table~\ref{tab:posteriors}.
The variance for half-lives that enters the likelihood also includes the contribution from the $Q_\beta$ correction $\sigma^2_{Q_\beta}$.

\begin{table*}[t]
\caption{\label{tab:posteriors}%
Summary of the calibration posteriors: medians with the 16\% and 84\% quantiles in parentheses.
$V_0$ and $C_1^s$ in MeV\,fm$^3$, $\sigma^{\rm HL}_{\rm th}$ in $\log_{10}(T_{1/2}/{\rm s})$, $\sigma^{\rm GT}_{\rm th}$ in MeV.
All calibrations include the test emulator errors and the $Q_\beta$ errors. The Set A rows use the
six GT nuclei of Set A, and all others the eight GT nuclei of Table~\ref{tab:pool_gt}.}
\begin{ruledtabular}
\begin{tabular}{lllcccc}
Functional & Set & Likelihood & $V_0$ & $C_1^s$ & $\sigma^{\rm HL}_{\rm th}$ & $\sigma^{\rm GT}_{\rm th}$ \\
\colrule
SkO$'$ & A & Student-$t$ & $-290$ $(-326,\,-225)$ & $116.7$ $(99.0,\,128.7)$ & $0.49$ $(0.32,\,0.71)$ & $1.34$ $(0.91,\,1.95)$ \\
 &  & Normal & $-277$ $(-325,\,-199)$ & $108.2$ $(91.1,\,122.4)$ & $0.46$ $(0.34,\,0.63)$ & $1.20$ $(0.86,\,1.69)$ \\
 & G10 & Student-$t$ & $-287$ $(-323,\,-226)$ & $118.3$ $(109.9,\,125.9)$ & $1.00$ $(0.64,\,1.47)$ & $1.11$ $(0.76,\,1.57)$ \\
 &  & Normal & $-286$ $(-330,\,-208)$ & $115.7$ $(106.2,\,124.9)$ & $1.15$ $(0.87,\,1.53)$ & $0.98$ $(0.72,\,1.34)$ \\
 & L10 & Student-$t$ & $-290$ $(-324,\,-235)$ & $118.5$ $(109.9,\,125.9)$ & $0.91$ $(0.58,\,1.35)$ & $1.10$ $(0.76,\,1.57)$ \\
 &  & Normal & $-290$ $(-332,\,-217)$ & $115.4$ $(105.6,\,124.6)$ & $1.09$ $(0.82,\,1.44)$ & $0.98$ $(0.72,\,1.34)$ \\
 & Full & Student-$t$ & $-306$ $(-326,\,-278)$ & $115.0$ $(104.5,\,122.7)$ & $0.48$ $(0.38,\,0.62)$ & $1.16$ $(0.80,\,1.67)$ \\
 &  & Normal & $-314$ $(-336,\,-284)$ & $109.4$ $(97.4,\,119.3)$ & $0.50$ $(0.41,\,0.62)$ & $1.04$ $(0.76,\,1.45)$ \\
\colrule
SLy4 & A & Student-$t$ & $-317$ $(-375,\,-224)$ & $93.8$ $(78.2,\,105.6)$ & $0.51$ $(0.34,\,0.75)$ & $0.98$ $(0.61,\,1.46)$ \\
 &  & Normal & $-299$ $(-370,\,-197)$ & $88.1$ $(73.3,\,100.2)$ & $0.47$ $(0.35,\,0.65)$ & $0.82$ $(0.54,\,1.21)$ \\
 & G10 & Student-$t$ & $-397$ $(-423,\,-279)$ & $100.7$ $(93.1,\,106.5)$ & $1.08$ $(0.69,\,1.59)$ & $0.70$ $(0.37,\,1.10)$ \\
 &  & Normal & $-407$ $(-426,\,-385)$ & $97.9$ $(89.5,\,105.6)$ & $1.16$ $(0.87,\,1.57)$ & $0.64$ $(0.40,\,0.95)$ \\
 & L10 & Student-$t$ & $-265$ $(-322,\,-197)$ & $97.9$ $(89.3,\,104.4)$ & $0.56$ $(0.30,\,0.90)$ & $0.74$ $(0.42,\,1.14)$ \\
 &  & Normal & $-335$ $(-386,\,-263)$ & $96.7$ $(88.0,\,104.5)$ & $1.05$ $(0.76,\,1.41)$ & $0.63$ $(0.39,\,0.94)$ \\
 & Full & Student-$t$ & $-278$ $(-332,\,-215)$ & $95.2$ $(84.1,\,103.0)$ & $0.46$ $(0.34,\,0.59)$ & $0.79$ $(0.45,\,1.21)$ \\
 &  & Normal & $-403$ $(-417,\,-390)$ & $98.6$ $(90.8,\,105.7)$ & $0.61$ $(0.51,\,0.73)$ & $0.63$ $(0.40,\,0.93)$ \\
\colrule
UNEDF1-HFB & A & Student-$t$ & $-280$ $(-328,\,-180)$ & $109.0$ $(95.8,\,118.5)$ & $0.49$ $(0.32,\,0.70)$ & $0.93$ $(0.55,\,1.42)$ \\
 &  & Normal & $-258$ $(-322,\,-143)$ & $107.5$ $(95.0,\,117.6)$ & $0.41$ $(0.29,\,0.56)$ & $0.74$ $(0.46,\,1.16)$ \\
 & G10 & Student-$t$ & $-263$ $(-294,\,-221)$ & $112.5$ $(106.9,\,117.3)$ & $0.28$ $(0.11,\,0.52)$ & $0.66$ $(0.37,\,1.03)$ \\
 &  & Normal & $-256$ $(-286,\,-214)$ & $111.5$ $(104.9,\,117.3)$ & $0.18$ $(0.06,\,0.36)$ & $0.57$ $(0.37,\,0.86)$ \\
 & L10 & & \multicolumn{4}{l}{same set as G10} \\
 & Full & Student-$t$ & $-271$ $(-305,\,-208)$ & $105.6$ $(79.2,\,113.5)$ & $0.31$ $(0.22,\,0.41)$ & $0.92$ $(0.51,\,1.83)$ \\
 &  & Normal & $-257$ $(-291,\,-197)$ & $101.1$ $(76.3,\,111.2)$ & $0.25$ $(0.18,\,0.32)$ & $0.83$ $(0.49,\,1.56)$ \\
\end{tabular}
\end{ruledtabular}
\end{table*}

\begin{figure}[t]
\includegraphics[width=\columnwidth]{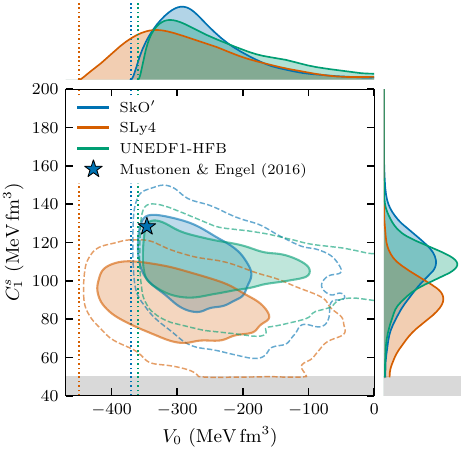}
\caption{\label{fig:posterior_A}%
Posterior distributions of $V_0$ and $C_1^s$ for Set A with the normal likelihood, for SkO$'$ (blue), SLy4 (orange), and UNEDF1-HFB (green).
The solid (dashed) contours enclose 68\% (95\%) of the posterior probability, and the marginal distributions are shown at the top and right of the panel.
The star marks the values of Ref.~\cite{MustonenEngel} for SkO$'$.
Dotted vertical lines mark $V_{\min}$ of each functional.
The shaded region, $C_1^s<50$~MeV\,fm$^3$, lies outside the stable domain.
}
\end{figure}

The resulting posterior distributions for the parameters $C^s_1$ and $V_0$ are shown in Fig.~\ref{fig:posterior_A}, and the summary statistics are shown in Table~\ref{tab:posteriors}.
The reported value from Ref.~\cite{MustonenEngel} of $(C^s_1, V_0) = (128.279, -346.352)$ for SkO$'$ ($V_0$ is listed there as $-173.176$, whereas in the current convention it is multiplied by a factor of 2) lies within the 68\% region of the posterior distribution in Fig.~\ref{fig:posterior_A}.
Note that in Ref.~\cite{MustonenEngel}, the weights entering $\chi^2$ were chosen by hand and the calculated $Q_\beta$ values were used without correction, which could contribute to the difference.

The posterior median of $C_1^s$ for SLy4 is about 20~MeV\,fm$^3$ lower than the other two.
The Landau--Migdal parameter $g_0'$ is linear in $C_1^s$, and a larger $g_0'$ leads to a larger GT resonance energy for a given single-particle spectrum~\cite{Bender2002}.
In our test with $C_1^s$ nearly switched off, the pnFAM GT resonance centroids of SLy4 lie above those of SkO$'$ and UNEDF1-HFB, though still below the measured values. 
Therefore, SLy4 does not require as large a $C_1^s$ as the other functionals to match the experiment. 
The posterior medians of the theory errors with the normal likelihood (Table~\ref{tab:posteriors}) are used to construct the information matrix for the E-optimal sets.

\subsection{E-optimal calibration sets \label{subsec:optimal_sets}}
Before identifying E-optimal sets, two nuclei per functional were removed from the pool because their calculated decay rates fall below the $10^{-9}$~s$^{-1}$ threshold of Sec.~\ref{subsec:training} over most of the parameter domain, leaving too few training samples for an emulator.
In total, the half-lives of 40 nuclei were included for each functional.
Tables~\ref{tab:pool_hl} and \ref{tab:pool_gt} show the complete list of nuclei included in the calibration.
All eight nuclei with measured GT resonances were included and entered the information matrix.

\begin{figure}[t]
\includegraphics[width=\columnwidth]{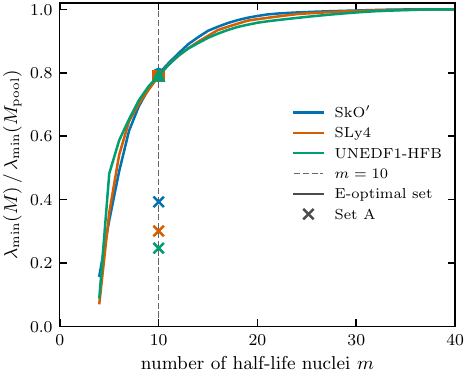}
\caption{\label{fig:information_saturation}%
Smallest eigenvalue of the information matrix, Eq.~\eqref{eq:mu}, of the local E-optimal set of $m$ half-life nuclei, with the sensitivities evaluated within the Set A posterior interval of $V_0$, Eq.~\eqref{eq:e-optimal}, normalized by the value for all 40 half-life nuclei.
All eight GT resonance energies are included at every $m$.
Filled symbols mark the L10 sets of Table~\ref{tab:calibsets}, and crosses mark Set A with the same eight GT resonance energies. 
}
\end{figure}

The sensitivity of the half-lives was evaluated globally across the training region and locally within the $V_0$ interval of the Set A calibration using the normal likelihood (Sec.~\ref{subsec:dataset} and Table~\ref{tab:posteriors}).
We refer to the sets of nuclei determined from the global and local sensitivity as the G10 and L10 sets, respectively, where G (L) stands for global (local) and 10 refers to the number of half-life nuclei, the same number as in Set A.
Figure~\ref{fig:information_saturation} shows how $\lambda_{\min}$ of Eq.~\eqref{eq:e-optimal} grows with the number $m$ of half-life nuclei. 
The L10 sets retain about 80\% of the information of the full pool of 40 half-lives.
This shows that the observables differ in how much they constrain the parameters, and the procedure provides a practical way to select the most informative ones.
We compare these E-optimal sets with the calibration performed with all 40 half-life nuclei, which we refer to as the Full set.
The resulting sets are summarized in Table~\ref{tab:calibsets}.
As shown in Fig.~\ref{fig:information_saturation}, the smallest eigenvalue of the information matrix for the L10 sets is two to three times that of Set A.
In contrast to Set A, by construction, each E-optimal set includes at least one nucleus with $A\geq200$.

\subsection{Calibrations with E-optimal sets \label{subsec:calib_optimal_sets}}

Initial Bayesian inferences of $\bm{\theta}$ have been performed with the G10, L10, and Full sets.
Based on the posterior distributions, we generate test samples to evaluate the final (test) error estimates of $\sigma_{\rm em}$ that enter the Bayesian inference.
For each functional, 40 test samples were drawn from a two-dimensional Latin hypercube~\cite{McKay1979} over $(V_0, C_1^s)$, bounded by the posterior medians $\pm$ three 68\% half-widths and by the prior support.
These bounds contain at least 97\% of the posterior distributions for the G10, L10, and Full sets (Table~\ref{tab:calibsets}) and for both likelihood functions.
The samples are then used in pnFAM to calculate each observable to be used in the emulator error evaluations.
The resulting test errors are shown in Fig.~\ref{fig:emulator_summary} and discussed in Sec.~\ref{subsec:emulator_performance}.

Using the test $\sigma_{\rm em}$, we perform Bayesian inference again to obtain the final posterior distributions.
The summary of the final posterior distributions is presented in Table~\ref{tab:posteriors}.
Focusing on the SLy4 functional, with the Student-$t$ likelihood, the L10 set gives a posterior more consistent with the Full set than the G10 set does.
For UNEDF1-HFB, the two sets are identical, and for SkO$'$ both give posteriors consistent with the Full set.
Therefore, in the subsequent discussions, we primarily focus on the L10 set.
For the choice of the likelihood function, we primarily focus on the Student-$t$ likelihood for both types of observables, as it is more robust to outliers.
The results with the normal likelihood are also listed for comparison.
For SLy4 with the Full set, the two likelihoods give different values of $V_0$.
This is because, with the normal likelihood, the posterior moves to reduce the large residuals of $^{52}$Ti and $^{74}$Zn, whereas the Student-$t$ likelihood reduces their weight.

\begin{figure*}[!htb]
\includegraphics[width=\textwidth]{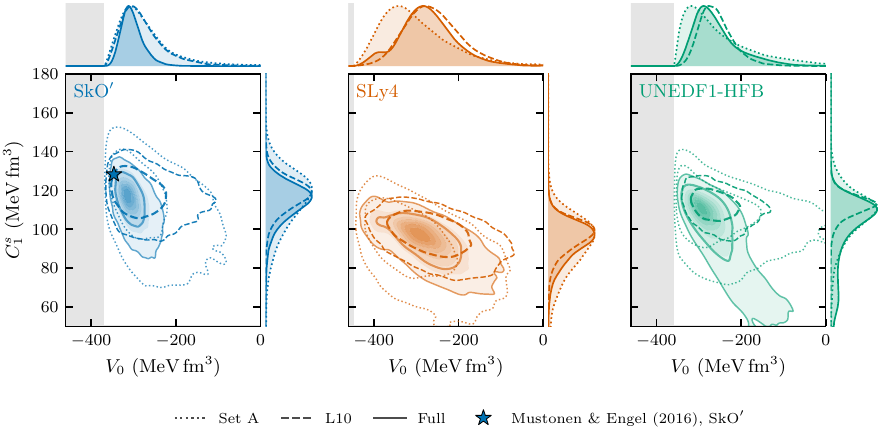}
\caption{\label{fig:posterior_final}%
Final posterior distributions of $V_0$ and $C_1^s$ with the Student-$t$ likelihood for each functional.
Filled contours show the Full set, dashed contours show the L10 set (identical to G10 for UNEDF1-HFB), and dotted contours show Set A. Inner and outer contours enclose 68\% and 95\% of the posterior probability, respectively.
The star marks the values of Ref.~\cite{MustonenEngel} for SkO$'$, and shaded regions lie outside the stable domain.
}
\end{figure*}

Figure~\ref{fig:posterior_final} shows the final posterior distributions obtained with the L10 and Full sets.
Compared with Set A, the E-optimal sets give intervals of $C_1^s$ narrower by about a factor of two for all three functionals, whereas the $V_0$ interval is narrower by a factor of two only for UNEDF1-HFB and by 10 to 20\% for SkO$'$ and SLy4 (Table~\ref{tab:posteriors}).
This shows that a larger information content does not guarantee a tighter posterior, partly because different nuclei prefer different parameter values. This tension is absorbed by the inferred theory errors and broadens the posterior, as seen for UNEDF1-HFB, where the Full set gives wider intervals than the L10 set.

\begin{figure*}[!htb]
\includegraphics[width=\textwidth]{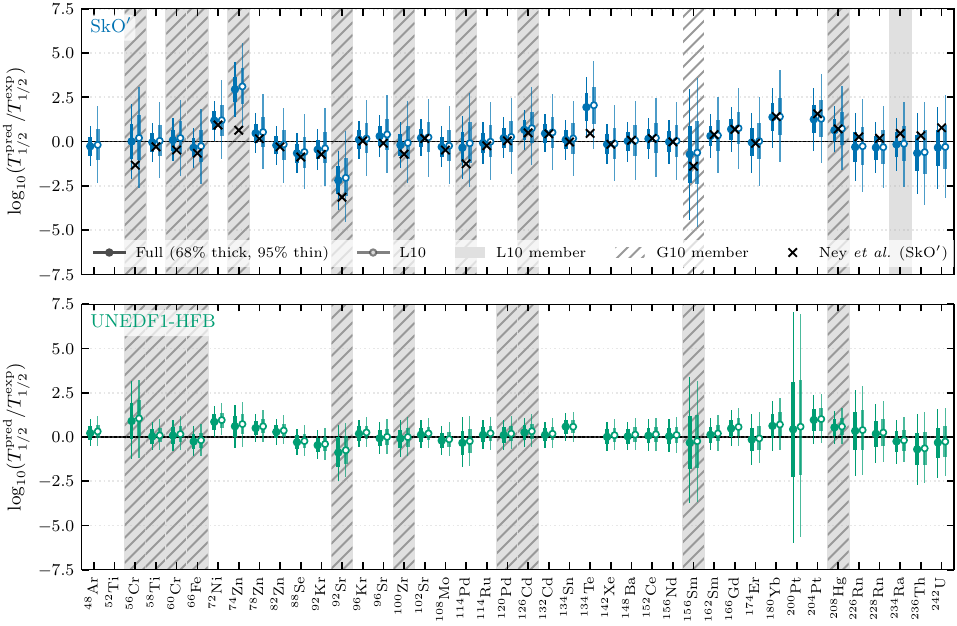}
\caption{\label{fig:pool_residuals}%
Posterior-predictive residuals for SkO$'$ and UNEDF1-HFB, $\log_{10}(T_{1/2}^{\rm calc}/T_{1/2}^{\rm exp})$, of the half-lives in the calibration pool with the Student-$t$ likelihood.
Results with SLy4 are similar to SkO$'$ and are omitted, but can be found in Ref.~\cite{calib_code}.
Filled symbols are from the Full-set calibration and open symbols from the L10 set.
Thick and thin bars are the 68\% and 95\% predictive intervals, respectively.
Gray (hatched) columns mark the members of the L10 (G10) sets, and crosses are the results of Ref.~\cite{Ney2020}.
Nuclei without symbols are not in the EDF calibration set  (Table~\ref{tab:pool_hl}).
}
\end{figure*}

The posteriors obtained with the Full set exhibit a stronger correlation between $V_0$ and $C_1^s$ for all functionals.
The origin is the following.
For nearly all nuclei, stronger $V_0$ and weaker $C_1^s$ move the half-lives in the same direction (shorten the half-lives).
The Full set contains four times as many half-lives as the L10 sets, so the half-lives contribute more to the likelihood.
For SkO$'$ and SLy4, this is reinforced by the smaller half-life theory error $\sigma^{\rm HL}_{\rm th}$ of the Full set (Table~\ref{tab:posteriors}), since the effect of the few nuclei with large residuals (see Fig.~\ref{fig:pool_residuals}) is diluted. This exposes the correlated structure of $V_0$ and $C_1^s$ due to half-lives.
In the case of UNEDF1-HFB, the half-lives additionally pull $C_1^s$ below the value preferred by the GT resonance centroid energies, which increases the $\sigma^{\rm GT}_{\rm th}$ and extends the posterior toward low $C_1^s$.

\begin{figure}[!htb]
\includegraphics[width=\columnwidth]{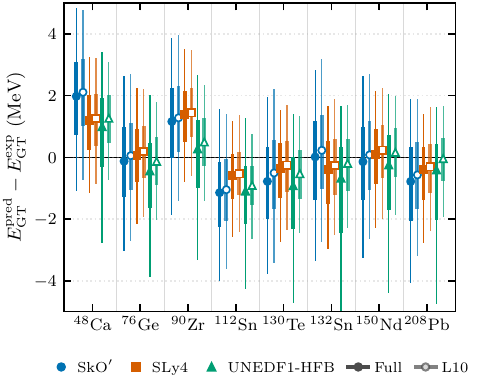}
\caption{\label{fig:gt_residuals}%
Posterior-predictive residuals of the eight GT resonance energies, with the Student-$t$ likelihood.
Filled (open) symbols are from the Full (L10) set.
Thick and thin bars are the 68\% and 95\% predictive intervals, respectively.
All eight nuclei are included in both calibrations.
}
\end{figure}

Figure~\ref{fig:pool_residuals} shows that the half-lives calibrated with the L10 and Full sets are largely consistent. 
This agreement supports the E-optimal construction of the calibration sets.
Since the L10 calibration predicts the half-lives not included in the set nearly as well as the Full calibration, the current calibration is expected to generalize reasonably well.
Compared to SkO$'$ and SLy4 (omitted in the figure but similar to SkO$'$), the residuals with UNEDF1-HFB remain small across the nuclei in the calibration pool.
The GT resonance centroid energies are also calibrated reasonably well (Fig.~\ref{fig:gt_residuals}).
However, since every nucleus shown in the figure is used in the calibration, how this performance generalizes to other nuclei cannot be inferred from it.

\subsection{Prediction of the half-lives of the \texorpdfstring{$N=126$}{N=126} isotonic chain \label{subsec:N126}}
To demonstrate the impact of our Bayesian calibration and uncertainty quantification method, we computed the $\beta$-decay half-lives of even-even nuclei along the $N=126$ isotonic chain.
The calculated $Q_\beta$ values before and after the linear correction are shown in Table~\ref{tab:n126_gs}.
As discussed in Sec.~\ref{subsec:bayes}, the half-life prediction $\hat{y}_k(\bm\theta)$ of each $N=126$ isotone is approximated by a quadratic surface in $(V_0, C_1^s)$ fitted to 20 pnFAM calculations at parameter values drawn from the Full-set posterior with the Student-$t$ likelihood.

\begin{figure}[!htb]
\includegraphics[width=\columnwidth]{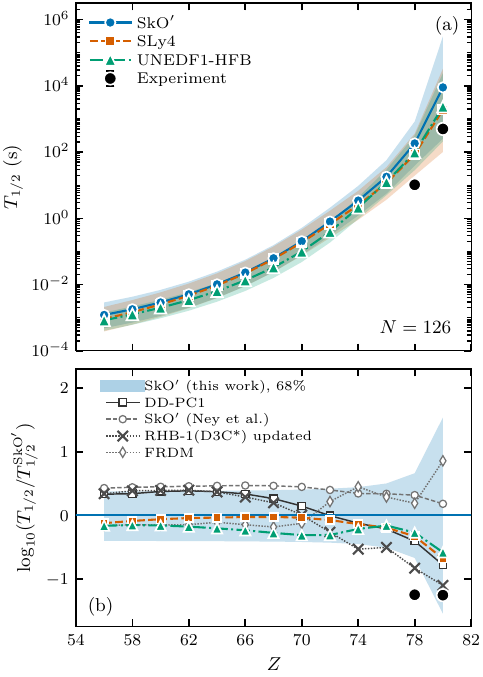}
\caption{\label{fig:n126}%
(a) Predicted half-lives of the even-even $N=126$ isotones from the Full-set calibrations with the Student-$t$ likelihood.
Lines with symbols are the half-lives at the posterior-median parameters, and the bands are the 68\% credible intervals.
Black circles are the measured half-lives of $^{204}$Pt and $^{206}$Hg~\cite{Kondev2021}.
(b) Half-lives relative to our SkO$'$ results at the posterior-median parameters: our SLy4 and UNEDF1-HFB results,  SkO$'$ results of Ref.~\cite{Ney2020}, DD-PC1 predictions~\cite{RSN2026}, the updated RHB-1(D3C*) half-lives reported in Ref.~\cite{Robin2024}, FRDM results~\cite{Moller2019}, and the measured values.
The shaded band is the SkO$'$ 68\% credible interval.
}
\end{figure}

Figure~\ref{fig:n126} shows the predicted half-lives with 68\% credible intervals and compares them with other global calculations~\cite{Ney2020,RSN2026,Moller2019,Robin2024}.
Experimental data are available for two isotones, $^{204}$Pt, which is part of the calibration, and $^{206}$Hg, which is not.
While the measured half-life of $^{206}$Hg lies within the 68\% intervals, the predictions for both nuclei are longer than the measured values.
Two effects contribute to the overestimate.
First, the corrected $Q_\beta$ values of these nuclei (Table~\ref{tab:n126_gs}) are 0.3 to 0.8~MeV below the experimental values, which results in the longer predicted half-lives.
Second, the half-lives of low-$Q_\beta$ nuclei are determined by the low-energy tail of the $\beta^-$ strength within the small $Q_\beta$ window.
The relativistic calculations of Ref.~\cite{Robin2024} reported that the coupling of quasiparticles to vibrations shifts strength to lower energies and shortens the half-lives for $^{204}$Pt and $^{206}$Hg; such correlations are not included in the present framework.

\begin{figure*}[!htb]
\includegraphics[width=\textwidth]{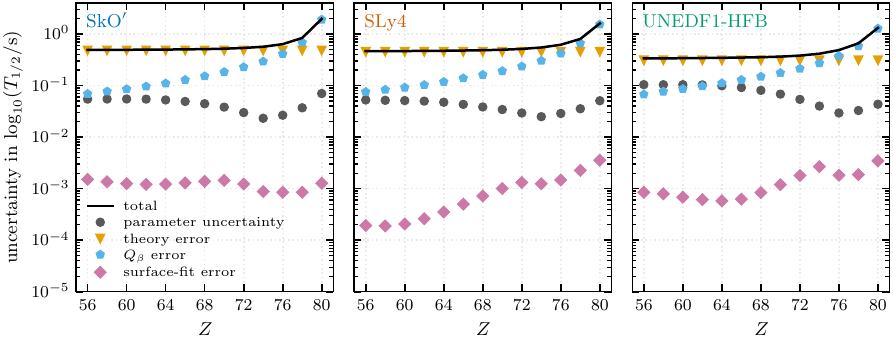}
\caption{\label{fig:n126_error}%
Contributions (standard deviations) to the computed uncertainty of the $N=126$ half-lives for each functional with the Full set and the Student-$t$ likelihood: parameter uncertainty, inferred theory error, propagated $Q_\beta$ error, and the error of the quadratic surface fitted to 20 pnFAM calculations, estimated with pnFAM calculations not used in the fit.
}
\end{figure*}

Figure~\ref{fig:n126_error} shows that the inferred theory error dominates the uncertainty budget up to $Z=76$ ($Z=74$ for UNEDF1-HFB), and the $Q_\beta$ uncertainty dominates for larger $Z$.
The parameter uncertainty is smaller than the theory error by a factor of 3 to 20, and the error introduced by the quadratic fit is negligible.

Considering the dominant contributions from the inferred theory error and the $Q_\beta$ uncertainty, reducing the uncertainty of the predictions for the $N=126$ isotones cannot be done by tightening constraints on $V_0$ and $C_1^s$.
The QRPA restricts the excitations of the daughter nucleus to two-quasiparticle configurations.
Possible improvements include coupling to higher-order configurations, through quasiparticle-vibration coupling~\cite{Niu2015,Robin2024,Liu2024} or four-quasiparticle configurations~\cite{Gambacurta2020}, both of which would fragment the strength and shift it down to lower energies.

For SkO$'$, our predictions give consistently shorter half-lives than those of Ref.~\cite{Ney2020}, which used the parameters of Ref.~\cite{MustonenEngel} and uncorrected $Q_\beta$ values.
Evaluated at the parameters of Ref.~\cite{MustonenEngel}, our SkO$'$ half-lives remain shorter by factors of 2--3 along the chain.
The remaining difference is therefore attributed mainly to the $Q_\beta$ correction.
Other  EDF-based results agree with each other below $Z=70$ and consistently predict longer half-lives than ours at the median parameter values (Fig.~\ref{fig:n126}(b)).

\begin{figure}[t]
\includegraphics[width=\columnwidth]{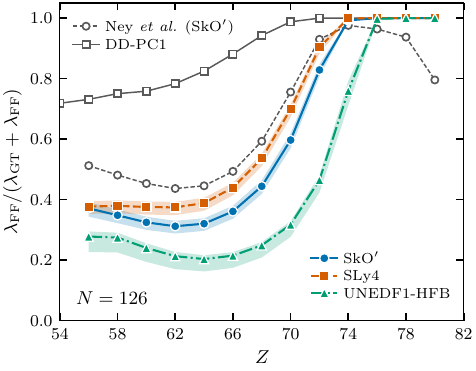}
\caption{\label{fig:n126_ff}%
Ratio of the first-forbidden decay rate to the total $\beta$-decay rate, $\lambda_{\rm FF}/(\lambda_{\rm GT}+\lambda_{\rm FF})$, of the even-even $N=126$ isotones from the Full-set calibrations with the Student-$t$ likelihood, compared with the SkO$'$ results of Ref.~\cite{Ney2020} and the DD-PC1 results of Ref.~\cite{RSN2026}.
Markers are the values at the posterior median, and the bands are the 16--84\% range from the parameter uncertainty only.
}
\end{figure}

As shown in Fig.~\ref{fig:n126_ff}, the contributions of the FF decay to the total decay rates vary significantly between the relativistic EDF framework (DD-PC1~\cite{Niksic2008,RSN2026}) and the nonrelativistic framework (SkO$'$, SLy4, and UNEDF1-HFB), as also noted in Ref.~\cite{RSN2026}.
The quantified uncertainty includes only the contribution from parameter uncertainties, as we do not have estimates of the theory error or contributions from $Q_\beta$ values.
In addition to the significant difference between the results of DD-PC1 and SkO$'$ of Ref.~\cite{Ney2020}, our results predict an even smaller contribution of FF decay for $Z\leq72$, with UNEDF1-HFB being the smallest.
Above $Z=74$, our calculations predict nearly pure FF decay, unlike the SkO$'$ tabulation of Ref.~\cite{Ney2020}.
Currently, the calibrations do not directly incorporate the information on FF decays.
Therefore, experimental constraints on the FF decays, obtained with, e.g., total absorption spectroscopy, may further test and constrain the EDF predictions.

In Ref.~\cite{RSN2026}, it has been pointed out that the half-lives of neutron-rich nuclei past $N=126$ predicted by the EDF-based approaches with axial deformation are longer than those of FRDM and RHB-1(D3C*) of Ref.~\cite{Marketin_RQRPA_2016}.
The longer half-lives were shown to reduce the flow of the neutron-induced fission in neutron-rich conditions such as disk ejecta from binary neutron star mergers, and hinder the production of the third $r$-process peak elements in magnetorotational supernovae.
It will be of future interest to investigate how our model, with quantified uncertainties, for the $N=126$ nuclei and beyond would affect the neutron-induced fission flow and the synthesis of third-$r$-process-peak nuclei ($A\sim195$), as well as actinides, under various astrophysical conditions.

\section{Conclusions \label{sec:conclusions}}
In this work, we have performed the Bayesian calibration of the isoscalar pairing strength $V_0$ and the spin density coupling constant $C_1^s$ in the proton-neutron FAM calculations with three energy density functionals, SkO$'$, SLy4, and UNEDF1-HFB, utilizing the recently developed linear response emulators.
Theoretical $Q_\beta$-values were finetuned to experimental values to improve the predictions of half-lives.
A comprehensive set of sources of uncertainty was considered in the Bayesian framework: theory error, $Q_\beta$-values, parameter uncertainty, and emulator errors.
Based on these calibrations, we also quantified uncertainties in $\beta$-decay half-lives for $N=126$ isotones.

We introduced an E-optimal experimental design for choosing the calibration targets, avoiding any parameter combination being poorly constrained.
The E-optimal sets of ten half-lives, combined with all eight GT resonance energies, retain about 80\% of the information of the full pool of 40 half-lives, measured by the smallest eigenvalue of the information matrix.
The predictions from the E-optimal L10 sets agree well with those from the Full set, implying that the calibrations would generalize reasonably well.
Among the three functionals, UNEDF1-HFB shows the best agreement with the measured half-lives from mass number $A=48$ to 242, while SLy4 exhibits the smallest theory error for the GT resonance centroid energies with the Full set.

To quantify the uncertainty, we computed the $\beta$-decay half-lives and credible intervals of the $N=126$ isotones with even proton numbers, relevant to the astrophysical $r$-process.
Over most of the $N=126$ chain, the predictive uncertainty is dominated by the theory error, while the uncertainty due to the $Q_\beta$ values dominates close to the stability line.
The uncertainty from the calibrated parameters is orders of magnitude smaller, suggesting that improvements in predictions should come from modeling $\beta$ decay and $Q_\beta$ values.
In this work, we treated $V_0$ as a constant; in the future, a more flexible treatment, such as a density-dependent $V_0$, may be explored.
Furthermore, expanding the calibration scope to include time-even terms may improve the accuracy of $Q_\beta$ predictions.

Toward the neutron-rich end of the isotonic $N=126$ chain, our predicted half-lives are shorter than those of previous EDF-based calculations.
The contribution of first-forbidden decay varies significantly between the relativistic and nonrelativistic frameworks, as well as between the nonrelativistic EDFs explored in this work.
Experimental data on the contribution of first-forbidden decay may further test and constrain the theoretical predictions.
While large-scale calculations are left for future work, if the predicted trend persists above $N=126$, the understanding of the production of the third $r$-process peak elements ($A\sim195$) and the roles of $\beta$-delayed and neutron-induced fission would be impacted.

\begin{acknowledgments}
We thank J. Engel (University of North Carolina at Chapel Hill) for providing access to the pnFAM and pynFAM codes and for helpful discussions.
We also thank K. Godbey (Facility for Rare Isotope Beams, Michigan State University) for helpful discussions.
This work was supported by the U.S. National Science Foundation under grant number 26-23480 (NP3M), the U.S. Department of Energy under Award No. DE-NA0004245 (NNSA, the Stewardship Science Academic Alliances program), and the DOE Office of Science under Grants DE-SC0013365 and DE-SC0023175 (Office of Advanced Scientific Computing Research and Office of Nuclear Physics, Scientific Discovery through Advanced Computing).
Computational resources were provided in part by the Institute for Cyber-Enabled Research at Michigan State University.
The authors used Claude Code (Anthropic), an AI coding assistant, to help develop and run the scripts for the emulator training, the Bayesian calibration, the data analysis, and the figures and tables, and to check and improve the manuscript text.
The authors reviewed the results and the text and take full responsibility for the content of this article.
\end{acknowledgments}

\section*{Data availability}
The trained emulators, the posterior samples of all calibrations, and the scripts that produce every figure and table of this article are openly available in the code repository~\cite{calib_code}. The pnFAM outputs from which the emulator training data were extracted are available in the companion data record~\cite{calib_data}.

\appendix
\section{Information matrix \label{app:information_matrix}}
The dependence of theoretical predictions on the parameters $\bm{\theta}=(V_0, C_1^s)$ is generally nonlinear.
Therefore, the  measured data can be expressed as
\begin{equation}
    y(x) = \eta(x, \bm{\theta}) + \varepsilon(x), \label{eq:nonlinear_model}
\end{equation}
where $x$ is the label for a nucleus and the type of observable, $y$ is the experimental value of the observable of interest for $x$, $\eta$ is the model prediction ($\beta$-decay logarithmic half-life $\log_{10}T_{1/2}$ or the Gamow--Teller resonance centroid energy), and $\varepsilon$ is the residual, containing experimental, theoretical, $Q_\beta$, and emulator contributions.
In the following, $y$ and $\varepsilon$ are the shorthand notations for $y(x)$ and $\varepsilon(x)$, respectively.
For the purpose of constructing the optimal design, let us assume that
\begin{align}
    \operatorname{E}[\varepsilon(x)]&=0, \nonumber \\
    \operatorname{Var}[\varepsilon(x)] &= \sigma^2(x), \nonumber \\
    \operatorname{Cov}[\varepsilon(x), \varepsilon(x')] &= 0 \quad(x \neq x'), \label{eq:design_assumption}
\end{align}
with $\sigma^2(x)$ given in Sec.~\ref{subsec:dataset}.

In terms of the scaled parameters $\bm{z}=(v,c)$ of Eq.~(\ref{eq:scaled_params}), a first-order Taylor expansion around a reference point $\bm{z}_t$ gives
\begin{equation}
    \eta(x, \bm{z}) \simeq \eta(x, \bm{z}_t) + (\bm{z} - \bm{z}_t)^T \bm{f} (x, \bm{z}_t), \label{eq:taylor}
\end{equation}
where
\begin{equation}
    \bm{f} (x, \bm{z}_t) = \left.\frac{\partial \eta(x, \bm{z})}{\partial \bm{z}}\right|_{\bm z=\bm z_t}, \label{eq:derivative}
\end{equation}
is the local sensitivity vector.
Note that $\bm{z}_t$ does not need to be a global optimum.
Combining with Eq.~\eqref{eq:nonlinear_model},
\begin{align}
    y(x) - \eta(x, \bm{z}_t) \simeq (\bm{z} - \bm{z}_t)^T \bm{f} (x, \bm{z}_t) + \varepsilon(x),
\end{align}
thus, the nonlinear calibration is approximated by a local linear regression.
Minimizing the objective
\begin{equation}
    \sum_{x\in \mathcal{C}}\frac{[y-\eta(x, \bm{z}_t)-(\bm{z} - \bm{z}_t)^T \bm{f}(x, \bm{z}_t)]^2}{\sigma^2(x)},
\end{equation}
for the calibration set $\mathcal{C}$, one obtains~\cite[Sec.~1.4.2, Eqs.~(1.72)--(1.76)]{fedorov2013optimal}
\begin{gather}
    \bm{M}_\mathcal{C}(\bm{z}_t)(\hat{\bm{z}}_\mathcal{C}-\bm{z}_t) \simeq \mathcal{Y}_\mathcal{C}(\bm{z}_t),
\end{gather}
where $\bm{M}_\mathcal{C}$ is the information matrix of Eq.~\eqref{eq:mu} and
\begin{equation}
    \mathcal{Y}_\mathcal{C}(\bm{z}_t) = \sum_{x \in \mathcal{C}}\frac{[y - \eta(x, \bm{z}_t)]\bm{f}(x, \bm{z}_t)}{\sigma^2(x)}.
\end{equation}

The inverse of $\bm{M}_\mathcal{C}$ approximates the 
variance matrix in the scaled parameter space:
\begin{equation}
    \operatorname{Var}[\hat{\bm z}_{\mathcal C}]
    \simeq \bm M_{\mathcal C}^{-1}.
\end{equation}
In the case of normally distributed observations, $\bm{\mu}(x, \bm{z}_t)$ is called the Fisher information matrix~\cite[Secs.~1.1.2 and 1.5]{fedorov2013optimal}.

\clearpage

\section{Nuclear data and ground-state properties \label{Sec:Appendix-data}}
Tables~\ref{tab:pool_hl} and \ref{tab:pool_gt} list the candidate nuclei for the calibration, and Table~\ref{tab:n126_gs} displays the $Q_\beta$ values of the $N=126$ isotones.

\begin{table*}[t]
\caption{\label{tab:pool_hl}%
Nuclei whose $\beta$-decay half-lives are part of the calibration pool, and their
experimental half-lives and uncertainties from Ref.~\cite{Kondev2021}, as well as the predicted quadrupole
deformation $\beta_2$ of the HFB ground state obtained with each functional (I=SkO', II=SLy4, III=UNEDF1-HFB).
A dagger marks the two nuclei per functional that are
excluded from that functional's pool because their decay rates fall below the resolution of the
calculation (Sec.~\ref{subsec:optimal_sets}).}
\begin{ruledtabular}
\begin{tabular}{lrrrrrrr}
Nucleus & $Z$ & $N$ & $A$ & $T_{1/2}^{\rm exp}$ (s) & 
$\beta_2$(I) & $\beta_2$(II) & $\beta_2$(III) \\
\colrule
$^{48}$Ar & 18 & 30 & 48 & $0.415 \pm 0.015$ & -0.16 & -0.12 & -0.16 \\
$^{52}$Ti & 22 & 30 & 52 & $102 \pm 6$ & 0.00$^\dagger$ & 0.00 & 0.00$^\dagger$ \\
$^{58}$Ti & 22 & 36 & 58 & $0.055 \pm 0.006$ & 0.00 & 0.00 & 0.00 \\
$^{56}$Cr & 24 & 32 & 56 & $356 \pm 6$ & +0.18 & 0.00$^\dagger$ & +0.16 \\
$^{60}$Cr & 24 & 36 & 60 & $0.490 \pm 0.010$ & 0.00 & 0.00 & -0.09 \\
$^{66}$Fe & 26 & 40 & 66 & $0.47 \pm 0.03$ & 0.00 & 0.00 & 0.00 \\
$^{72}$Ni & 28 & 44 & 72 & $1.57 \pm 0.05$ & 0.00 & 0.00 & 0.00 \\
$^{74}$Zn & 30 & 44 & 74 & $95.6 \pm 1.2$ & 0.00 & 0.00 & +0.16 \\
$^{78}$Zn & 30 & 48 & 78 & $1.47 \pm 0.15$ & +0.17 & 0.00 & +0.14 \\
$^{82}$Zn & 30 & 52 & 82 & $0.178 \pm 0.003$ & +0.19 & +0.12 & +0.13 \\
$^{88}$Se & 34 & 54 & 88 & $1.53 \pm 0.06$ & -0.19 & +0.16 & +0.17 \\
$^{92}$Kr & 36 & 56 & 92 & $1.840 \pm 0.008$ & -0.21 & -0.16 & -0.15 \\
$^{96}$Kr & 36 & 60 & 96 & $0.080 \pm 0.008$ & -0.25 & -0.23 & -0.27 \\
$^{92}$Sr & 38 & 54 & 92 & $9400 \pm 61$ & -0.13 & -0.08 & -0.06 \\
$^{96}$Sr & 38 & 58 & 96 & $1.059 \pm 0.008$ & -0.22 & -0.17 & -0.13 \\
$^{102}$Sr & 38 & 64 & 102 & $0.069 \pm 0.006$ & +0.39 & +0.39 & +0.40 \\
$^{100}$Zr & 40 & 60 & 100 & $7.1 \pm 0.4$ & -0.20 & -0.19 & -0.19 \\
$^{108}$Mo & 42 & 66 & 108 & $1.105 \pm 0.010$ & -0.20 & -0.20 & -0.21 \\
$^{114}$Ru & 44 & 70 & 114 & $0.54 \pm 0.03$ & 0.00 & 0.00 & -0.21 \\
$^{114}$Pd & 46 & 68 & 114 & $145 \pm 4$ & 0.00 & +0.09 & +0.20 \\
$^{120}$Pd & 46 & 74 & 120 & $0.49 \pm 0.03$ & 0.00 & 0.00 & 0.00 \\
$^{126}$Cd & 48 & 78 & 126 & $0.512 \pm 0.005$ & 0.00 & 0.00 & 0.00 \\
$^{132}$Cd & 48 & 84 & 132 & $0.084 \pm 0.005$ & 0.00 & 0.00 & 0.00 \\
$^{134}$Sn & 50 & 84 & 134 & $0.93 \pm 0.08$ & 0.00 & 0.00 & 0.00 \\
$^{134}$Te & 52 & 82 & 134 & $2508 \pm 48$ & 0.00 & 0.00$^\dagger$ & 0.00$^\dagger$ \\
$^{142}$Xe & 54 & 88 & 142 & $1.23 \pm 0.02$ & +0.17 & +0.15 & +0.16 \\
$^{148}$Ba & 56 & 92 & 148 & $0.620 \pm 0.005$ & +0.23 & +0.22 & +0.22 \\
$^{152}$Ce & 58 & 94 & 152 & $1.42 \pm 0.02$ & +0.30 & +0.29 & +0.30 \\
$^{156}$Nd & 60 & 96 & 156 & $5.06 \pm 0.13$ & +0.34 & +0.33 & +0.34 \\
$^{156}$Sm & 62 & 94 & 156 & $(3.38 \pm 0.07)\times10^{4}$ & +0.33 & +0.32 & +0.32 \\
$^{162}$Sm & 62 & 100 & 162 & $2.7 \pm 0.3$ & +0.34 & +0.34 & +0.33 \\
$^{166}$Gd & 64 & 102 & 166 & $5.1 \pm 0.8$ & +0.34 & +0.33 & +0.33 \\
$^{174}$Er & 68 & 106 & 174 & $192 \pm 12$ & +0.31 & +0.30 & +0.31 \\
$^{180}$Yb & 70 & 110 & 180 & $144 \pm 30$ & +0.29 & +0.28 & +0.29 \\
$^{200}$Pt & 78 & 122 & 200 & $(4.54 \pm 0.11)\times10^{4}$ & 0.00$^\dagger$ & -0.08 & -0.08 \\
$^{204}$Pt & 78 & 126 & 204 & $10.3 \pm 1.4$ & 0.00 & 0.00 & 0.00 \\
$^{208}$Hg & 80 & 128 & 208 & $135 \pm 10$ & 0.00 & 0.00 & 0.00 \\
$^{226}$Rn & 86 & 140 & 226 & $444 \pm 6$ & +0.16 & +0.17 & 0.00 \\
$^{228}$Rn & 86 & 142 & 228 & $65 \pm 2$ & +0.19 & +0.19 & 0.00 \\
$^{234}$Ra & 88 & 146 & 234 & $30 \pm 10$ & +0.24 & +0.24 & +0.24 \\
$^{236}$Th & 90 & 146 & 236 & $2238 \pm 90$ & +0.26 & +0.25 & +0.25 \\
$^{242}$U & 92 & 150 & 242 & $1008 \pm 30$ & +0.27 & +0.26 & +0.26 \\
\end{tabular}
\end{ruledtabular}
\end{table*}

\begin{table*}[t]
\caption{\label{tab:pool_gt}%
Nuclei whose Gamow--Teller strength distributions enter the calibration,
with the experimental Gamow--Teller resonance peak energies, their uncertainties, and their sources.
Predicted $\beta_2$ as in Table~\ref{tab:pool_hl}.}
\begin{ruledtabular}
\begin{tabular}{lrrrrcrrr}
Nucleus & $Z$ & $N$ & $A$ & $E_{\rm GT}^{\rm exp}$ (MeV) & Ref. & $\beta_2$(I) & $\beta_2$(II) & $\beta_2$(III) \\
\colrule
$^{48}$Ca & 20 & 28 & 48 & $10.0 \pm 0.5$ & \cite{Ca48GT} & 0.00 & 0.00 & 0.00 \\
$^{76}$Ge & 32 & 44 & 76 & $11.1 \pm 0.5$ & \cite{Ge76Te130GT} & +0.14 & 0.00 & +0.16 \\
$^{90}$Zr & 40 & 50 & 90 & $9.0 \pm 0.5$ & \cite{Zr90GT} & 0.00 & 0.00 & 0.00 \\
$^{112}$Sn & 50 & 62 & 112 & $8.9 \pm 0.25$ & \cite{Sn112GT} & 0.00 & 0.00 & 0.00 \\
$^{132}$Sn & 50 & 82 & 132 & $16.3 \pm 0.5$ & \cite{Sn132GT} & 0.00 & 0.00 & 0.00 \\
$^{130}$Te & 52 & 78 & 130 & $13.6 \pm 0.5$ & \cite{Ge76Te130GT} & 0.00 & 0.00 & 0.00 \\
$^{150}$Nd & 60 & 90 & 150 & $15.2 \pm 0.5$ & \cite{Nd150GT} & +0.31 & +0.24 & +0.26 \\
$^{208}$Pb & 82 & 126 & 208 & $15.6 \pm 0.2$ & \cite{Pb208GT} & 0.00 & 0.00 & 0.00 \\
\end{tabular}
\end{ruledtabular}
\end{table*}

\begin{table*}[t]
\caption{\label{tab:n126_gs}%
HFB ground-state properties of the even-even $N=126$ isotones with each functional: 
the $\beta$-decay $Q$ value before ($Q_\beta^{\rm HFB}$) and after ($Q_\beta^{\rm corr}$) the correction of Sec.~\ref{subsec:qcorr}, in MeV.
The experimental values for $^{202}$Os and $^{206}$Hg are from 
AME2020~\cite{AME2020} (\# marks the value from systematics)
and for $^{204}$Pt from  Ref.~\cite{Freire2026}.
All ground states are predicted to be spherical.}
\begin{ruledtabular}
\begin{tabular}{lrrrrrrrr}
 & & \multicolumn{2}{c}{SkO$'$} & \multicolumn{2}{c}{SLy4} & \multicolumn{2}{c}{UNEDF1-HFB} & Exp. \\
Nucleus & $Z$ & $Q_\beta^{\rm HFB}$ & $Q_\beta^{\rm corr}$ & $Q_\beta^{\rm HFB}$ & $Q_\beta^{\rm corr}$ & $Q_\beta^{\rm HFB}$ & $Q_\beta^{\rm corr}$ & $Q_\beta^{\rm exp}$ \\
\colrule
$^{182}$Ba & 56 & 16.02 & 18.27 & 16.92 & 17.81 & 17.52 & 18.18 & -- \\
$^{184}$Ce & 58 & 14.27 & 16.35 & 15.33 & 16.14 & 15.44 & 15.97 & -- \\
$^{186}$Nd & 60 & 12.87 & 14.77 & 13.97 & 14.68 & 13.85 & 14.25 & -- \\
$^{188}$Sm & 62 & 11.39 & 13.12 & 12.46 & 13.08 & 12.34 & 12.61 & -- \\
$^{190}$Gd & 64 & 9.89 & 11.44 & 10.87 & 11.39 & 10.87 & 11.01 & -- \\
$^{192}$Dy & 66 & 8.38 & 9.76 & 9.24 & 9.67 & 9.48 & 9.49 & -- \\
$^{194}$Er & 68 & 7.09 & 8.29 & 7.98 & 8.32 & 8.31 & 8.19 & -- \\
$^{196}$Yb & 70 & 5.84 & 6.87 & 6.77 & 7.02 & 7.21 & 6.96 & -- \\
$^{198}$Hf & 72 & 4.67 & 5.52 & 5.57 & 5.72 & 6.11 & 5.74 & -- \\
$^{200}$W & 74 & 3.62 & 4.30 & 4.37 & 4.43 & 5.01 & 4.51 & -- \\
$^{202}$Os & 76 & 2.55 & 3.06 & 3.24 & 3.21 & 3.91 & 3.27 & $4.11 \pm 0.50$\# \\
$^{204}$Pt & 78 & 1.51 & 1.84 & 2.16 & 2.03 & 2.87 & 2.10 & $2.60 \pm 0.01$ \\
$^{206}$Hg & 80 & 0.51 & 0.66 & 1.10 & 0.88 & 1.85 & 0.96 & $1.31 \pm 0.02$ \\
\end{tabular}
\end{ruledtabular}
\end{table*}

\bibliography{bibl}

\end{document}